\documentclass[sigconf, nonacm]{acmart}

\newcommand{\va}{\vspace{-.5mm}}

\usepackage[ruled,vlined,linesnumbered]{algorithm2e}

\begin{document}
\title{Empowering Investigative Journalism with Graph-based Heterogeneous Data Management}

%%
%% The "author" command and its associated commands are used to define the authors and their affiliations.
\author{Angelos-Christos Anadiotis}
\affiliation{%
  \institution{\'Ecole Polytechnique, IPP \& EPFL}
}
\email{angelos.anadiotis@polytechnique.edu}

\author{Oana Balalau}
% \author{Th\'eo Bouganim}
% \author{Francesco Chimienti}
\affiliation{%
  \institution{Inria \& IPP}
}
\email{oana.balalau@inria.fr}

\author{Th\'eo Bouganim}
\affiliation{%
  \institution{Inria \& IPP}
}
\email{theo.bouganim@inria.fr}

\author{Francesco Chimienti}
\affiliation{%
  \institution{Inria \& IPP}
}
\email{francesco.chimienti@inria.fr}

\author{Helena Galhardas}
\affiliation{%
  \institution{INESC-ID \& IST, Univ. Lisboa}
}
\email{hig@inesc-id.pt}

\author{Mhd~Yamen~Haddad}
\affiliation{%
  \institution{Inria \& IPP}
}
\email{mhd-yamen.haddad@inria.fr}

\author{St\'ephane Horel}
\affiliation{%
  \institution{Le Monde}
}
\email{horel@lemonde.fr}

\author{Ioana Manolescu}
% \author{Youssr~Youssef}
\affiliation{%
  \institution{Inria \& IPP}
}
\email{ioana.manolescu@inria.fr}

\author{Youssr~Youssef}
\affiliation{%
  \institution{Inria \& IPP}
}
\email{youssr.youssef@inria.fr}

% \author{Angelos-Christos~Anadiotis$^{\star}$, Oana~Balalau$^{\diamond}$, Th\'eo Bouganim$^{\diamond}$, Francesco~Chimienti$^{\diamond}$, Helena~Galhardas$^{\ddagger}$, Mhd~Yamen~Haddad$^{\diamond}$,  St\'ephane~Horel$^{\dagger}$, Ioana~Manolescu$^{\diamond}$, Youssr~Youssef$^{\diamond}$}
% \affiliation{%
%   \institution{$^{\star}$\'Ecole Polytechnique,~IPP~\&~EPFL,
%                 $^{\diamond}$Inria~\&~IPP,
%                 $^{\ddagger}$INESC-ID~\&~IST,~Univ.~Lisboa,
%                 $^{\dagger}$Le~Monde
%               }
% }
% \email{name.surname@{polytechnique.edu, inria.fr}, hig@inesc-id.pt, horel@lemonde.fr}
\begin{abstract}
Investigative Journalism (IJ, in short) is  staple of modern,
democratic societies.  % and may play an important role in the shaping of public opinion and society-level decisions.
IJ often necessitates
working with {\em large, dynamic sets of heterogeneous, schema-less data sources}, which can be
{\em structured, semi-structured, or textual}, limiting the applicability of
classical data integration approaches.
In prior work, we have developed ConnectionLens, a system capable of integrating such sources into a single heterogeneous graph, leveraging Information Extraction (IE) techniques; users can then query the graph by means of keywords, and explore query results and their neighborhood using an interactive GUI. Our keyword search problem is complicated by the graph heterogeneity, and by the lack of a result score function that would allow to prune some of the search space.

In this work, we describe an {\em actual IJ application} studying conflicts of interest in the biomedical domain, and we show how ConnectionLens supports it. Then, we present {\em novel techniques addressing the scalability challenges} raised by this application: one allows to reduce the significant IE costs while building the graph, while the other is a novel, parallel, in-memory keyword search engine, which achieves orders of magnitude speed-up over our previous engine. Our experimental study on the real-world IJ application data confirms the benefits of our contributions.
%This work is a collaboration between CS researchers and an
%investigative journalist. We present an original IJ application
%suggested by the journalist, and show how to implement it using the ConnectionLens system we have been developing. Our algorithmic
%contributions are
\end{abstract}

\maketitle

% \pagestyle{\vldbpagestyle}
% \begingroup\small\noindent\raggedright\textbf{PVLDB Reference Format:}\\
% \vldbauthors. \vldbtitle. PVLDB, \vldbvolume(\vldbissue): \vldbpages, \vldbyear.\\
% \href{https://doi.org/\vldbdoi}{doi:\vldbdoi}
% \endgroup
% \begingroup
% \renewcommand\thefootnote{}\footnote{\noindent
% This work is licensed under the Creative Commons BY-NC-ND 4.0 International License. Visit \url{https://creativecommons.org/licenses/by-nc-nd/4.0/} to view a copy of this license. For any use beyond those covered by this license, obtain permission by emailing \href{mailto:info@vldb.org}{info@vldb.org}. Copyright is held by the owner/author(s). Publication rights licensed to the VLDB Endowment. \\
% \raggedright Proceedings of the VLDB Endowment, Vol. \vldbvolume, No. \vldbissue\ %
% ISSN 2150-8097. \\
% \href{https://doi.org/\vldbdoi}{doi:\vldbdoi} \\
% }\addtocounter{footnote}{-1}\endgroup

% \ifdefempty{\vldbavailabilityurl}{}{
% \vspace{.3cm}
% \begingroup\small\noindent\raggedright\textbf{PVLDB Artifact Availability:}\\
% The source code, data, and/or other artifacts have been made available at \url{\vldbavailabilityurl}.
% \endgroup

\newcommand\mysection[1]{\vspace{-2mm}\section{#1}\vspace{-1mm}}
\newcommand\mysubsection[1]{\vspace{-1mm}\subsection{#1}\vspace{-.5mm}}

\mysection{Introduction}
\label{sec:introduction}

Journalism and the press are a critical ingredient of any modern society. Like many other industries, such as trade, or entertainment, journalism has benefitted from the explosion of Web technologies, which enabled instant sharing of their content with the audience. However, unlike trade, where databases and data warehouses had taken over daily operations decades before the Web age, {\em many newsrooms discovered the Web and social media, long before building strong information systems where journalists could store their information and/or ingest data of interest for them.} As a matter of fact, journalists' desire to protect their confidential information may also have played a role in delaying the adoption of data management infrastructures in newsrooms.

At the same time, highly appreciated journalism work often requires {\em acquiring, curating, and exploiting large amounts of digital data}.
Among the authors, S.~Horel co-authored the ``Monsanto Papers'' series which obtained the European Press Prize Investigative Reporting Award in 2018~\cite{monsanto}; a similar project is the ``Panama Papers'' (later known as ``Offshore Leaks'') series of the International Consortium of Investigative Journalists~\cite{panama}.
In such works, journalists are forced to work with {\em heterogeneous} data, potentially in {\em different data models} ({\em structured} such as relations, {\em semistructured} such as JSON or XML documents, or graphs, including but not limited to RDF, as well as {\em unstructured text}).
We, the authors, are currently collaborating on such an \textbf{Investigative Journalism} (IJ, in short) application, focused on the study of \textbf{situations potentially leading to conflicts of interest}\footnote{According to the 2011 French transparency law, ``A conflict of interest is any situation where a public interest may interfere with a public or private interest, in such a way that the public interest may be, or appear to be, unduly influenced.''} (CoIs, in short) between biomedical experts and various organizations: 
corporations, industry associations, lobbying organizations or front groups. 
Information of interest in this setting comes from: scientific publications (in PDF) where authors declare e.g., ``Dr. X. Y. has received consulting fees from ABC''; semi-structured metadata (typically XML, used for instance in \href{https://pubmed.ncbi.nlm.nih.gov/}{PubMed}), where authors may also specify such connections; a medical association, say, French cardiology, may build its own disclosure database which may be relational, while a company may  disclose its ties to specialists in a spreadsheet.

This paper builds upon our recent work~\cite{DBLP:journals/corr/abs-2012-08830}, where we have identified a set of requirements (\textbf{R}) and the constraints (\textbf{C}) that need to be addressed to efficiently support IJ applications. 
%The solution we advocate is {\em integrating all sources into a single, heterogeneous graph}. , and 
We recall them here for clarity and completeness: 

\noindent\textbf{R1. Integral source preservation and provenance:} in journalistic work, it is crucial to be able to trace each information item back to the data source from which it came. This enables {\em adequately sourcing} information, an important tenet of quality journalism.

\noindent\textbf{R2. Little to no effort required from users:} journalists often lack time and resources to set up IT tools or data processing pipelines. Even when they are able to use a tool supporting one or two data models (e.g., most relational databases provide some support for JSON data), handling other data models remains challenging.
Thus, the data analysis pipeline needs to be as automatic as possible.

\noindent\textbf{C1. Little-known entities:} interesting journalistic datasets feature some extremely well-known entities (e.g., world leaders in the pharmaceutical industry) next to others of much smaller notoriety (e.g., an expert consulted by EU institutions, or a little-known trade association). From a journalistic perspective, such lesser-known entities may play a crucial role in making interesting connections among data sources, e.g., the association may be created by the industry leader,  and it may pay the expert honoraries. 

\noindent\textbf{C2. Controlled dataset ingestion:} the level of confidence in the data required for journalistic use excludes massive ingestion from uncontrolled data sources, e.g., through large-scale Web crawls.

%\noindent\textbf{C3. Language support:} journalists are first and foremost concerned with the affairs surrounding them (at the local or national scale). This requires supporting dataset in the language(s) relevant for them - in our case, French.\ACA{Do we want to keep this constraint here?}

\noindent\textbf{R3. Performance on ``off-the-shelf'' hardware:} The efficiency of our data processing pipeline is important; also,  the tool should run on general-purpose hardware, available to users like the ones we consider, without expertise or access to special hardware. %\IM{I removed 'experts' because Stéphane is an expert in her own right :D OK ,an expert in something else... and when I tried to add "IT expert", it took another line...}

Further, IJ applications' data analysis needs entail: 

\noindent\textbf{R4. Finding connections across heterogeneous datasets} is a core need. In particular, it is important for our approach to be tolerant of inevitable  differences in the organization of data across sources. Heterogeneous data integration works, such as~\cite{DBLP:books/daglib/0029346,DBLP:conf/dlog/CalvaneseGLLPR07,DBLP:journals/pvldb/BuronGMM20}, and recent heterogeneous polystores, e.g.,~\cite{DugganESBHKMMMZ15,DBLP:conf/sigmod/AlotaibiBDMZ19,DBLP:conf/cidr/QuamarST20} assume that sources have well-understood schemas; other recent works, e.g.,~\cite{DBLP:journals/pvldb/OtaMFS20,DBLP:journals/pvldb/Christodoulakis20,DBLP:conf/sigmod/NargesianPZBM20} focus on analyzing large sets of Open Data sources,  all of which are tabular. IJ data sources do not fit these hypothesis: data can be semi-structured, structured, or simply text. Therefore, we opt for \textbf{integrating all data sources in a heterogeneous graph} (with no integrated schema), and for \textbf{keyword-based querying} where users specify some terms, and the system returns subtrees of the graph, that connect nodes matching these terms. 

\noindent\textbf{C4. Lack of single, well-behaved answer score:} After discussing several journalistic scenarios, no unique method (score) for deciding which are the best answers to a query has been identified. Instead: ($i$)~it appears that ``very large'' answers (say, of more than 20 edges) are of limited interest; ($ii$)~connections that ``state the obvious'', e.g., that a French scientist is connected to a French company through their nationality, are not of interest. Therefore, unlike prior keyword search algorithms, which fix a score function and exploit it to prune the search, our algorithm must be orthogonal and work it with any score function.

Building upon our previous work, and years-long discussions of IJ scenarios, this paper makes the following contributions:

\va
\begin{itemize}
\item We describe the CoI IJ application proposed by S.~Horel (Section~\ref{sec:motivation}), we extract its technical requirements and we devise an end-to-end data analysis pipeline addressing these requirements (Section~\ref{sec:preliminaries}). 
\va
\item We provide application-driven optimizations, inspired from the CoI scenario but reusable to other contexts, which speeds up the graph construction process (Section~\ref{sec:optims}).
\va
\item We introduce a parallel, in-memory version of the keyword search algorithm previously introduced in~\cite{DBLP:journals/corr/abs-2009-04283, DBLP:journals/corr/abs-2012-08830}, and we explain our design in both the physical database layout and the parallel query execution (Section~\ref{sec:system}). 
\va
\item We evaluate the performance of our system using both synthetic and real-world PubMed data, we demonstrate its scalability, and we show that we have improved the performance compared to our prior work by several orders of magnitude, thereby enabling the journalists to perform interactive exploration of their data (Section~\ref{sec:evaluation}). 
\va 
\end{itemize}

%This paper is organized as follows.
%Section~\ref{sec:motivation} presents the CoI investigation we collaborate on. Then, Section~\ref{sec:preliminaries} presents the IJ pipeline we have built, and the previous capabilities of our system. Then, Section~\ref{sec:optims} describes a scenario-inspired optimization which speeds up graph constructions, while
%Section~\ref{sec:system} describes a novel keyword search algorithm, leveraging parallelism and in-memory speed for maximum efficiency. Section~\ref{sec:evaluation} presents our experimental evaluation; we then discuss related works and conclude.

\mysection{Use case: conflicts of interest in the biomedical domain}
\label{sec:motivation}

\noindent\textbf{The topic.}
Biomedical experts such as health scientists and researchers in life sciences play an important role in society, advising governments and the public on health issues. They also routinely interact with industry (pharmaceutical, agrifood etc.), consulting, collaborating on research, or otherwise sharing work and interests. To trust advice coming from these experts, it is important to ensure  the advice is not unduly influenced by vested interests. Yet, IJ work, e.g.~\cite{oreskes,lobbytomie,nicotine}, has shown that disclosure information is often scattered across multiple data sources, hindering access to this information.  We now illustrate the data processing required to gather and collectively exploit such information.

\begin{figure}[t!]
\includegraphics[width=\columnwidth]{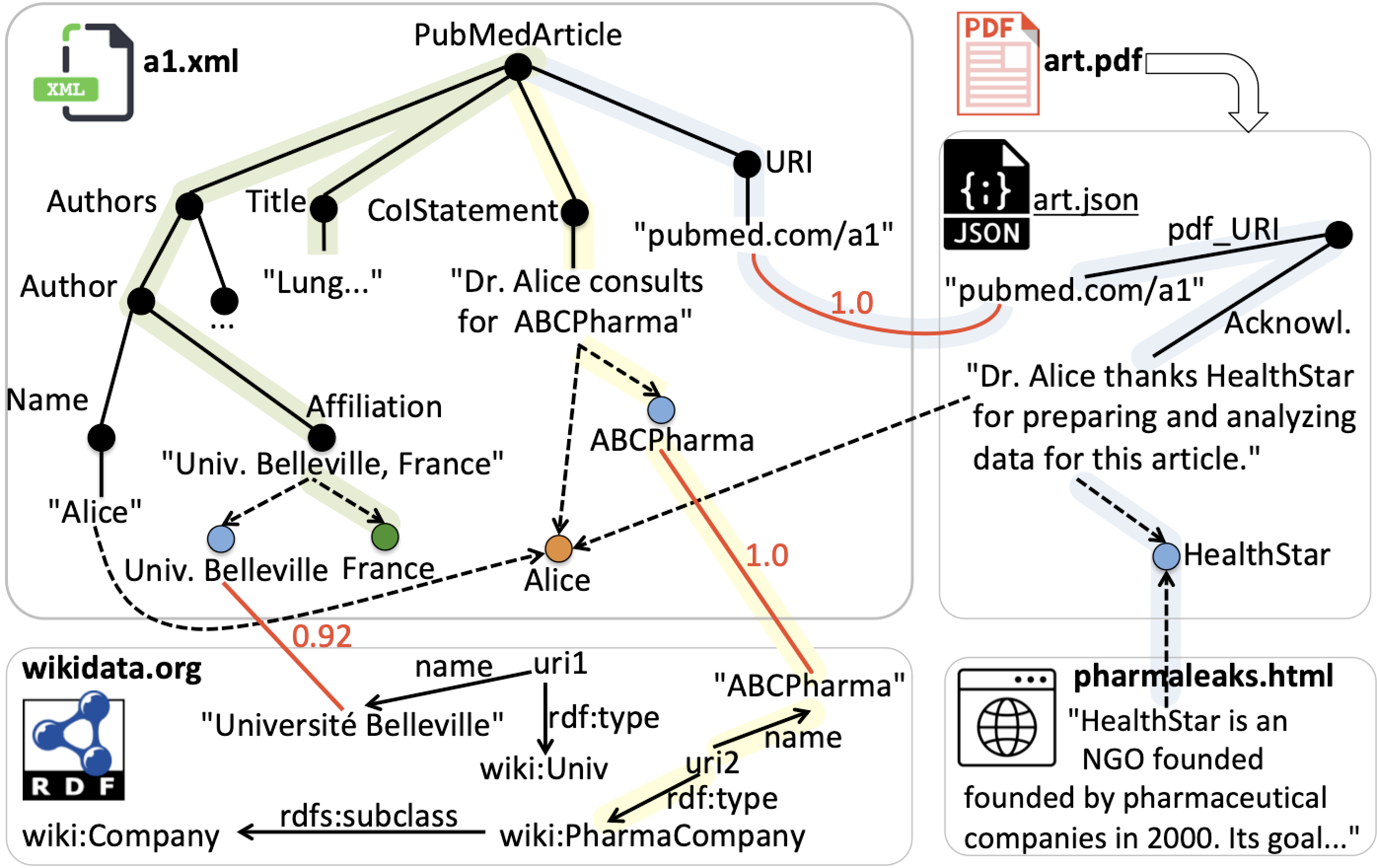}
\vspace{-6mm}
\caption{Graph data integration in ConnectionLens.\label{fig:motiv}}
\vspace{-6mm}
\end{figure}

\noindent\textbf{Sample data.} Figure~\ref{fig:motiv} shows a tiny fragment of data that can be used to find connections between scientists and companies. {\em For now, consider only the nodes shown as a black dot or as a text label, and the solid, black edges connecting them; these model directly the data. The others are added by ConnectionLens as we discuss in Section~\ref{sec:build-graph}.}
 ($i$)~Hundreds of millions of bibliographic notices (in \textbf{XML}) are published on the \href{https://pubmed.ncbi.nlm.nih.gov/}{PubMed} web site; the site also links to research (in PDF). In recent years, PubMed has included an optional CoIStatement element where authors can declare (in free text) their possible links with industrial players; less than 20\% of recent papers have this element, and some of those present, are empty (``The authors declare no conflict of interest''). ($ii$)~Within the \textbf{PDF} papers themselves,  paragraphs titled, e.g., ``Acknowledgments'', ``Disclosure statement'' etc. may contain such information, even if the CoIStatement is absent or empty. This information is accessible if one converts the PDF in a format such as \textbf{JSON}. In Figure~\ref{fig:motiv}, Alice declares her consulting for ABCPharma in XML, yet the ``Acknowledgments'' paragraph in her PDF paper mentions HealthStar\footnote{This example is inspired from prior  work of S.~Horel where she identified (manually inspecting thousands of documents) an expert supposedly with no industrial ties, yet who authored papers for which companies had supplied and prepared data.}. 
($iii$)~A (subset of a) knowledge base (in \textbf{RDF}) such as WikiData describes well-known entities, e.g., ABCPharma; however,  less-known entities of interest in an IJ scenario are often missing from such KGs, e.g., HealthStar in our example.  ($iv$)~Specialized data sources, such as a trade catalog or a Wiki Web site built by other investigative journalists, may provide information on some such actors: in our example, the PharmaLeaks Web site shows that HealthStar is also funded by the industry.  Such a site, established by a trusted source (or colleague), even if it has little or no structure, is a gold mine to be reused, since it saves days or weeks of tedious IJ work. {\em In this and many IJ scenarios, sources are highly heterogeneous, while  time, skills, and resources to curate, clean, or structure the data are not available}.

\noindent\textbf{Sample query.} Our application requires {\em the connections of specialists in lung diseases, working in France, with pharmaceutical companies}. In  Figure~\ref{fig:motiv}, the edges with {\em green} highlight and those with {\em yellow} highlight, together, form an answer connecting Alice to ABCPharma (spanning over the XML and RDF sources); similarly, the edges highlighted in {\em green} together with those in {\em blue}, spanning over XML, JSON and HTML,  connect her to HealthStar. 
 
\noindent\textbf{The potential impact of a CoI database.}  A database of known relationships between experts and interested companies, built by integrating heterogeneous data sources, would be a very valuable asset. In Europe, such a database could be used, e.g., to select, for a committee advising EU officials on %endocrine disruptors or 
industrial pollutants, experts with few or no such relationships. In the US, the Sunshine Act~\cite{sunshine}, just the French 2011 law, require manufacturers of drugs and medical devices to declare such information, but this does not extend to companies from other sectors.

\mysection{Investigative Journalism pipeline}
\label{sec:preliminaries}

\begin{figure}[h!]
\vspace{-2mm}
\includegraphics[width=\columnwidth]{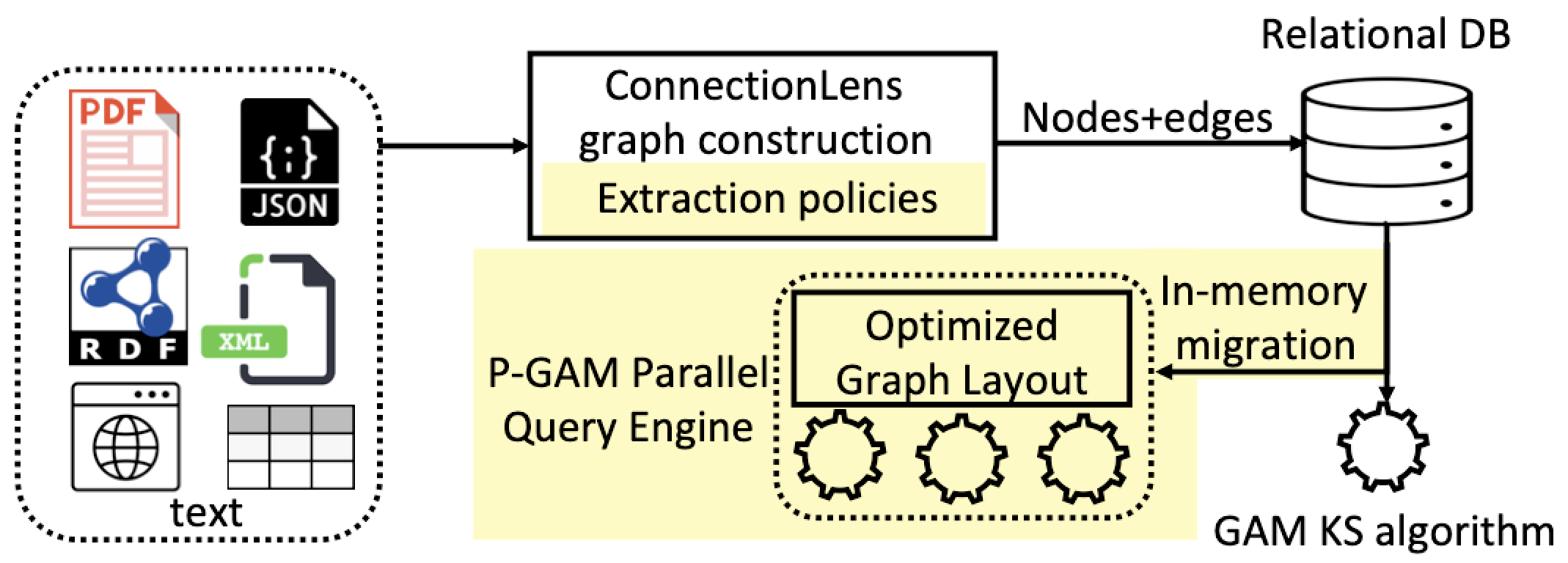}
\vspace{-8mm}
\caption{Investigative Journalism data analysis pipeline.\label{fig:archi}}
\vspace{-4mm}
\end{figure}

The  pipeline we have built for IJ is outlined in Figure~\ref{fig:archi}.  First, we recall ConnectionLens graph construction  (Section~\ref{sec:build-graph}), which  integrates heterogeneous data into a graph, stored and indexed in PostgreSQL. On this graph, the GAM keyword search algorithm (recalled in Section~\ref{sec:gam}) answers queries such as our motivating example; these are both detailed in~\cite{DBLP:journals/corr/abs-2012-08830}. The modules on yellow background in Figure~\ref{fig:archi} are the novelties of this work, and will be introduced below: scenario-driven performance optimizations to the graph construction (Section~\ref{sec:optims}), and an in-memory, parallel keyword search algorithm, called P-GAM (Section~\ref{sec:system}).

\mysubsection{ConnectionLens graph construction}
\label{sec:build-graph}
ConnectionLens integrates JSON, XML, RDF, HTML,  relational or text data  into a graph, as illustrated in Figure~\ref{fig:motiv}. 
%All edges shown in solid black lines derive directly from the structure of the XML, JSON, and RDF data, respectively. \IM{Already said earlier}
Each source is mapped to the graph as close to its data model as possible, e.g., XML edges have no labels while internal nodes all have names, while in JSON conventions are different etc.  Next, ConnectionLens   \textbf{extracts named entities from all text nodes}, regardless of the data source they come from,  using trained language models. In the figure, blue, green, and orange nodes denote Organization, Location, and Person entities, respectively. Each such entity node is connected to the text node it has been extracted from, by an {\em extraction edge} recording also the confidence of the extraction (dashed in the figure). \textbf{Entity nodes are shared across the graph}, e.g., Person:Alice has been found in three data sources, Org:BestPharma in two sources etc. ConnectionLens includes a {\em disambiguation} module which avoids mistakenly unifying entities with the same labels but different meanings. Finally, nodes with similar labels are {\em compared}, and if their similarity is above a threshold, a \textbf{sameAs} (red) edge is introduced connecting them, labeled with the similarity value.

A sameAs edge with similarity 1.0 is called an {\em equivalence edge}. Then, $p$  equivalent nodes, e.g., the ABCPharma entity and the identical-label RDF literal, would lead to $p(p-1)/2$ equivalence edges. To keep the graph compact, one of the $p$ nodes is declared the {\em representative} of all $p$ nodes, and instead, we only store the $p-1$ equivalence edges adjacent to the representative.
Details on all the above graph construction steps can be found in~\cite{DBLP:journals/corr/abs-2012-08830}.

Formally,  a ConnectionLens graph is denoted $G=(N,E)$, where nodes can be of different types (URIs, XML elements, JSON nodes etc., but also extracted entities) and edges encode: data source structure, entities extracted from text, and node label similarity.

\mysubsection{The GAM keyword search algorithm}
\label{sec:gam}
We view our motivating query, on highly heterogeneous content with no a-priori known structure, as a \textbf{keyword search query over a graph}. Formally, a query $Q=\{w_1, w_2, \ldots, w_m\}$ is a set of $m$ keywords, and an {\em answer tree} (AT, in short) is a set $t$ of $G$ edges which ($i$)~together, form a tree, and ($ii$)~for each $w_i$, contain at least one node whose label matches $w_i$.
We are interested in {\em minimal} answer trees, that is answer trees which satisfy the following properties: ($i$)~removing an edge from the tree will make it lack at least one keyword match, and ($ii$)~if more than one nodes match a query keyword, then all matching nodes are related through sameAs links with similarity 1.0.
In the literature (see Section~\ref{sec:related}), a {\em score function} is used to compute the quality of an answer, and only the best $k$ ATs are returned, for a small integer $k$.
Our problem is harder since: ($i$)~our ATs may span over different data sources, even of different data models; ($ii$)~they may traverse an edge \textbf{in its original or in the opposite direction}, e.g., to go from JSON to XML through Alice; this brings the search space size in $O(2^{|E|})$, where $|E|$ is the number of edges; and ($iii$)~\textbf{no single score function serves all IJ needs} since, depending on the scenario, journalists may favor different (incompatible) properties of an AT, such as ``being characteristic of the dataset'' or, on the contrary, ``being surprising''. Thus, \textbf{we cannot rely on special properties of the score function}, to help us prune unpromising parts of the search space, as done in prior work (see Section~\ref{sec:related}). 
Intuitively,  tree size could be used to limit the search: very large answer trees (say, of more than 100 edges) generally do not represent meaningful connections. However, in heterogeneous, complex graphs, users find it hard to set a size limit for the exploration. Nor is a smaller solution always better than a larger one. For instance, an expert and a company may both have ``nationality'' edges leading to ``French'' (a solution of 2 edges), but that may be less interesting than finding that the expert has written an article specifying in its CoIStatement funding from the company (which could span over 5 edges or more).

\begin{figure}
\includegraphics[width=.9\columnwidth]{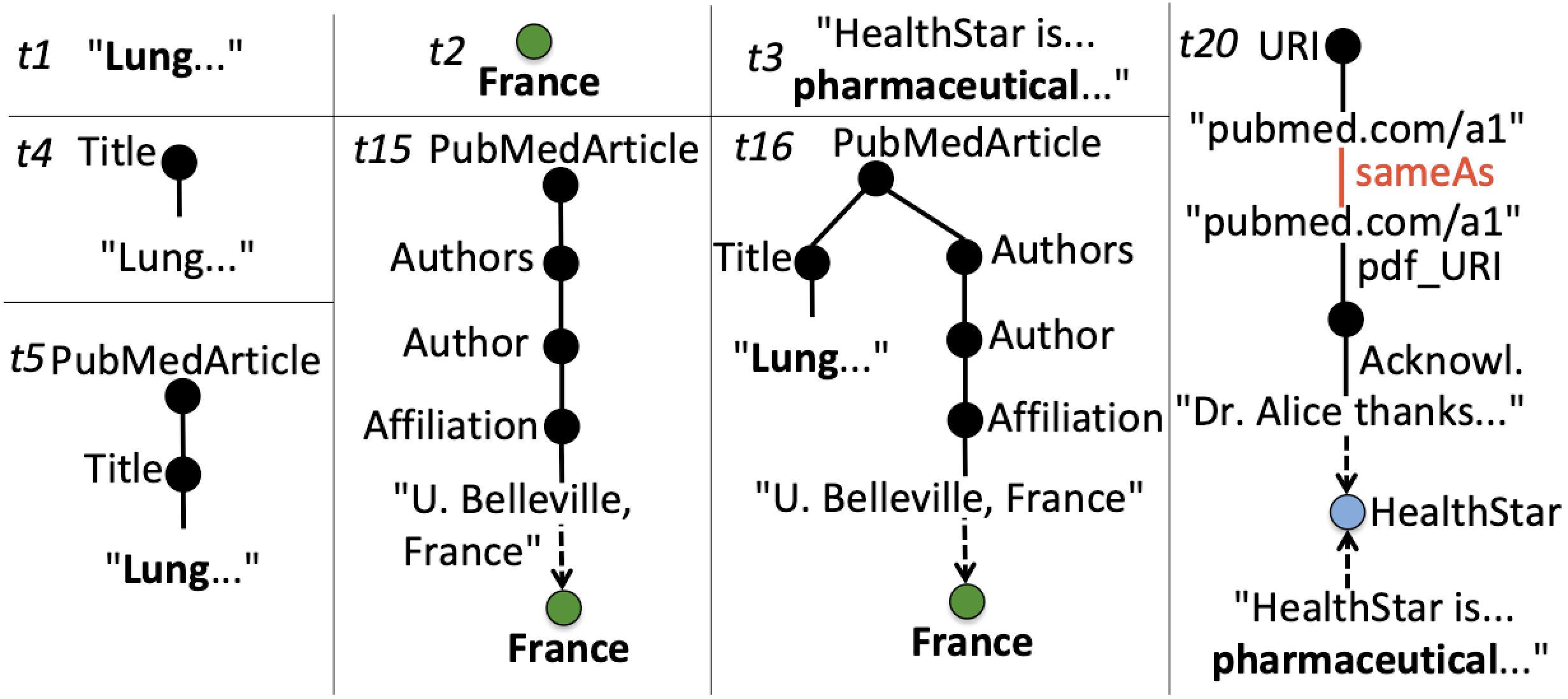}
\vspace{-3mm}
\caption{Trees built by GAM for our sample query.\label{fig:GAM-trees}}
\vspace{-6mm}
\end{figure}

Our  \textbf{Grow-and-Aggressive-Merge (GAM)} algorithm~\cite{DBLP:journals/corr/abs-2009-04283,DBLP:journals/corr/abs-2012-08830} enumerates trees exhaustively, until a number of answers are found, or a time-out. First, it builds 1-node trees from the nodes of $G$ which match 1 or more keywords, e.g., $t_1,t_2,t_3$ in Figure~\ref{fig:GAM-trees}, showing some partial trees built when answering our sample query. The keyword match in each node label appears in bold.
Then, GAM relies on two steps. \textbf{Grow} adds to the root of a tree one of its adjacent edges in the graph, leading to a new tree: thus $t_4$ is obtained by Grow on $t_1$, $t_5$ by Grow on $t_4$, and successive Grow steps lead from $t_2$ to $t_{15}$. Similarly, from $t_3$, successive Grow's go from the HTML to the JSON data source (the HealthStar entity occurs in both), and then to the XML one, building $t_{20}$. Second, as soon as a tree is built by Grow, it is \textbf{Merge}d with all the trees already found, rooted in the same node, matching different keywords and having disjoint edges wrt the given tree.  For instance, assuming $t_{15}$ is built after $t_5$, they are immediately merged into the tree $t_{16}$, having the union of their edges. Each Merge result is then merged again with all qualifying trees (thus the ``agressive'' in the algorithm name). For instance, when Grow on $t_{20}$ builds a tree rooted in the PubMedArticle node (not shown; call it $t_A$),  Merge($t_{16},t_A$) is immediately built, and is exactly the answer highlighted with green and blue in Figure~\ref{fig:motiv}.

Together, Grow and Merge are guaranteed to generate all solutions. If $m=2$, Grow alone is sufficient, while $m\leq 3$ requires also the Merge step. {\em GAM may build a tree in several ways}, e.g., the answer above could also be obtained as Merge(Merge($t_{15}$, Grow($t_{20}$)), $t_5$); GAM keeps a history of the trees already explored, to avoid repeating work on them.
Importantly, GAM can be used with any score function. Its details are described in \cite{DBLP:journals/corr/abs-2009-04283,DBLP:journals/corr/abs-2012-08830}.

%Equivalence edges are explored by Grow as follows: it can only {\em go from one node, to its representative} (if they differ); if a solution combines 3 or more equivalent nodes,
%which GAM could assemble in $2^{p(p-1)/2}$ ways; many are useless, since equivalence is transitive.
%and Grow is only allowed to traverse an equivalence edge {\em to go from one node, to its representative} (if they differ). This preserves search completeness while avoiding some useless trees.

%\IM{Need to figure out if ``tricks other than the parallel GAM''
%  deserve a separate section? Maybe ``optimization and extensions''?}
\mysection{Use case-driven optimization}
\label{sec:optims}
In this section, we present an optimization we brought to the graph construction process, guided by our target application.

In the experiments we ran, Named Entity Recognition (NER) took up to 90\% of the time ConnectionLens needs to integrate data sources into a graph.
The more textual the sources are, the more time is spent on NER. %; conversely, if the data contains few text nodes, the NER effort is lowered. 
Our application data lead us to observe that:

\begin{itemize}
\item Some text nodes, e.g., those found on the path PubMedArticle.Authors.Author.Name,  {\em always correspond to entities of a certain type}, in our example, Person. If this information is given to ConnectionLens, it can create a Person entity node, like the Alice node in Figure~\ref{fig:motiv}, {\em without calling the expensive NER procedure}. 
\item Other text nodes may be deemed {\em uninteresting for the extraction},  journalists think no interesting entities appear  there. If ConnectionLens is aware of this, it can {\em skip the NER call on such text nodes}. Observe that the input data, including all its text nodes, is always preserved; we only avoid extraction effort deemed useless (but which can still be applied later if application requirements evolve).
\end{itemize}

To exploit this insight, we introduced a notion of \textbf{context}, and allow users to specify \textbf{(optional) extraction policies}. A context is an expression designating a set of text nodes in one or several data sources. For instance, a context specified by the rooted path PubMedArticle.Authors.Author.Name designates  all the text values of nodes found on that path in an XML data source; the same mechanism applies to an HTML or JSON data source. 
In a relational data source containing table $R$ with attribute $a$, a context of the form  $R.a$ designates all text nodes in the ConnectionLens graph obtained from a value of the attribute $a$ in relation $R$. Finally, an RDF property $p$ used as context designates all the values $o$ such that a triple $(s,p,o)$ is ingested in a ConnectionLens graph.

Based on contexts, an extraction policy takes one of the following form: ($i$) \underline{$c$ {\em force} $T_e$} where $c$ is a context and $T_e$ is an entity type, e.g., Person, states that each node designated by the context is exactly one instance of $T_e$ ; ($ii$) \underline{$c$ {\em skip}}, to indicate that NER should not be performed on the text nodes designated by $c$; ($iii$)~as syntactic sugar, for hierarchical data models (e.g., XML, JSON etc.),  \underline{$c$ {\em skipAll}} allows stating that NER should not be performed on the text nodes designated by $c$, nor on any descendant of their parent. This allows larger-granularity control of NER on different portions of the data.

% Contexts could easily be generalized to more powerful queries; we could also define them in text sources using patterns, e.g., ``any phrase starting with...''. 
Observe that our contexts (thus, our policies) are specified {\em within a data model}; this is because the {\em regularity that allows defining then} can only be hoped for within data sources with identical structure. %NER on all meaningful text nodes (e.g., not numbers or dates) remains the default; 
Policies allow journalists to state what is obvious to them, and/or what is not interesting, in the interest of graph construction speed. {\em Force} policies may also improve graph quality, by making sure NER does not miss any entity designated by the context. 

%\section{System Design and Implementation}
\mysection{In-memory parallel keyword search}
\label{sec:system}

We now describe the novel keyword search module that is the main technical contribution of this work.
A  {\em in-memory graph storage model} specifically designed for our graphs and with keyword search in mind (Section~\ref{sec:datamodel}) is leveraged by a 
 a {\em multi-threaded, paralell} algorithm, called P-GAM (Section~\ref{sec:parallelquery}), and which is  a parallel extension of our original GAM algorithm, outlined in Section~\ref{sec:gam}.

\mysubsection{Physical in-memory database design}
\label{sec:datamodel}

The size of the main memory in modern servers has grown significantly over the past decade.
For instance, AWS EC2 offers nodes providing up to 24TB of main memory and 448 hardware threads~\cite{awsexample}.
Data management research has by now led to several mature products (DB engines) running entirely in main memory,  such as Oracle Database In-Memory, SAP HANA, and Microsoft SQL Server with Hekaton.
Moving the data from the hard disk to the main memory  significantly boosts performance, avoiding disk I/O costs.
However, it introduces new challenges on the optimization of the data structures and the execution model for a different bottleneck: the memory access~\cite{10.5555/645925.671364}.
% In fact, query execution of in-memory databases has further pushed the shift in storage layout from to row to column stores~\cite{10.5555/1083592.1083658} as well as the execution model towards block- and vector-oriented~\cite{DBLP:conf/cidr/BonczZN05}.

% Drawing inspiration from the developments in column-stores,
We have integrated P-GAM inside a novel in-memory graph database, which we have built and optimized for P-GAM operations.
The physical layout of a graph database is important, given that graph processing is known to suffer from random memory accesses~\cite{DBLP:conf/isca/AhnHYMC15, DBLP:conf/fast/ElyasiCS19, DBLP:conf/sosp/RoyMZ13, DBLP:conf/sc/HongDMLVC15}.
Our design ($i$)~includes all the data needed by appplications as described in Section~\ref{sec:motivation}, while also ($ii$)~aiming at high performance,  parallel query execution in modern scale-up servers, in order to tackle huge search spaces (Section~\ref{sec:gam}).

%Due to its in-memory execution, the data layout plays a critical role in the performance of the search, since the performance of graph processing algorithms is well-known to suffer from random accesses.
%In our design, we consider two main requirements: (i) high performance in parallel query execution, and (ii) availability of all information required by the application.
%The first requirement stems from the need to achieve scalability in modern scale-up servers.
%The second requirement stems from the need to support the application scenario that we have already described in Section~\ref{sec:motivation}.

We start with the scalability requirements. Like GAM, P-GAM also performs Grow %, Grow2Rep, \IM{Trying to hide G2R under Grow, they are very similar anyway}
and Merge operations (recall Figure~\ref{fig:GAM-trees}).

To enumerate possible Grow steps, P-GAM needs to access all edges adjacent to the root of a tree, as well as the representative (Section~\ref{sec:build-graph}) of the root, to enable growing with an equivalence edge.
Further, as we will see, P-GAM (as well as GAM) relies on a simple edge metric, called \textbf{specificity}, derived from the number of edges with the same label adjacent to a given node, to decide the best neighbor to Grow to.
For instance, if a node has 1 spouse and 10 friend edges, the edge going to the spouse is more specific than one going to a friend.

%To grow a tree by adding it one edge adjacent to the tree root,  Grow requires to know the neighbors of every node as well as the specificity of every edge connecting the nodes, in order to apply the score function and select the best node to grow the intermediate answer tree.
%Further, GrowToRep needs to know the representative of every node, if there exists one.
A Merge does not need more information than available in its input trees; instead, it requires specific run-time data structures, as we describe below.

\setlength{\textfloatsep}{0pt}% Remove \textfloatsep

\begin{figure}
  \centering
  \includegraphics[width=.9\columnwidth]{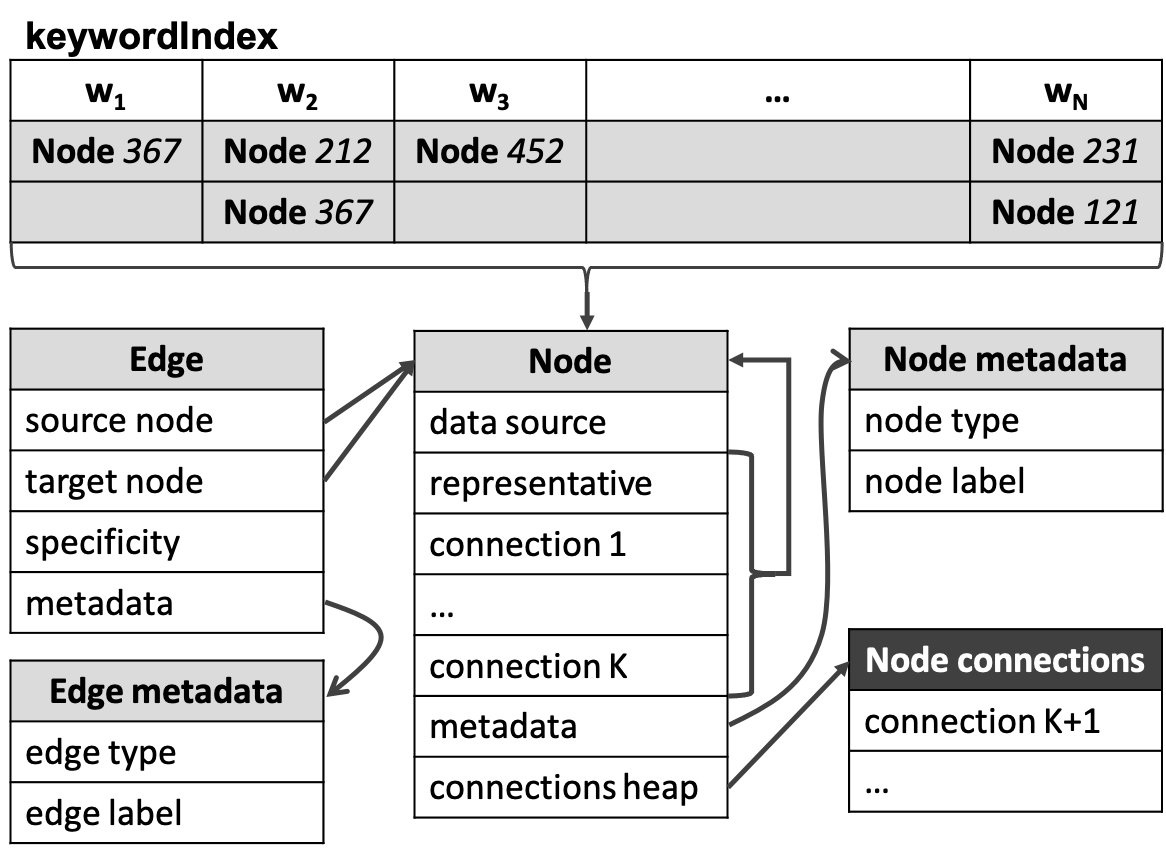}
\vspace{-4.5mm}
  \caption{Physical graph layout in memory.}
%\vspace{-7mm}
  \label{fig:dblayout}
\end{figure}

In our memory layout, we \textbf{split the data required for search, from the rest}, as the former are critical for performance; we refer to the latter as metadata.
Figure~\ref{fig:dblayout} depicts the memory tables that we use.
%\HG{with different colour nodes for the table names}. \IM{I just change the alignment, I think it's sufficient}
The {\em Node} table includes the ID of the data source where the node comes from, and  references to each node's: ($i$)~representative, %(if there exists one),
($ii$)~$K$ neighbors, if they exist (for a fixed $K$ - static allocation), ($iii$)~metadata, and ($iv$)~other neighbors, if they exist (dynamic allocation).
We separate the allocation of neighbors into static and dynamic, to keep  $K$ neighbors in the main Node structure, while the rest are placed in a separate heap area, stored in the {\em Node connections} table.
This way, we can allocate a fixed size to each Node, efficiently supporting the memory accesses of P-GAM.
In our implementation, we set $K=5$; in general, it can be set based on the average degree of the graph vertices.
The {\em Node metadata} table includes information about the type of each node (e.g., JSON, HTML, etc.) and its label, comprising the keywords that we use for searching the graph.
The {\em Edge} table includes a reference to the source and the target node of every edge, the edge specificity, and a reference to the edge metadata.
The {\em Edge metadata} table includes the type and the label of each edge.
Finally, we use a {\em keywordIndex}, which is a hash-based map associating every node with its labels.
P-GAM probes the keywordIndex when a query arrives to find the references to the Node table that match the query keywords and start the search from there.
Among all the structures, only {\em Node connections} (singled out by a dark background in Figure~\ref{fig:dblayout}) is in a dynamically allocated area; all the others are statically allocated.

The above storage is {\em row} (node) oriented, even though  column storage often speeds up greatly analytical processing; this is due to the nature of the keyword search problem, which requires traversing the graph from the nodes matching the keywords, in BFS style. % it accesses  neighbors of every visited node, to enumerate and prioritize (order) possible Grow steps.
Since we consider fully ad-hoc queries (any keyword combinations), there are no guarantees about the order of the nodes  P-GAM visits.
Therefore, in our setting, the vertically selective access patterns, which are optimally exploited by column-stores, do not apply.
Instead, the crucial optimization here is to {\em find the neighbors of every node fast.}
This is leveraged by our algorithm, as we explain below.
%Accordingly, our in-memory graph database engine implements a tuple-at-a-time model on top of a row-store \HG{maybe add a sentence explaining a bit more this idea}.\IM{I thought that while it was true that we are tuple-at-a-time, it does not strictly need to be said here.}

% \IM{To be inserted in this section, since it is built before querying:} a \textbf{keywordIndex} is built (as an array)  associating, to each query kwd$_i$, the nodes (specifically, the node positions in their statically allocated area, as described above) whose labels match kwd$_i$.

\begin{algorithm}[t]
\SetAlgoLined
\KwIn{$G=(N,E)$, query $Q$=\{w$_1$, $\ldots$, w$_m$\}, maximum number of solutions $M$, maximum time limit}
\KwOut{Answer trees for $Q$ on $G$}
%history $\leftarrow$ $\emptyset$; solutions $\leftarrow$ $\emptyset$\;
%$nt\leftarrow$ max($m$, max number of supported threads)\;
pQueue$_i$ $\leftarrow$ new priority queue of (tree, edge) pairs,  $1\leq i \leq nt$\;
$N_Q\leftarrow \cup_{w_i\in Q}$ keywordIndex.lookup(w$_i$)\;
\For{$n\in N_Q,e$ edge adjacent to $n$}{
  push $(n,e)$ on some pQueue$_j$ (distribute equally)
}
launch $nt$ P-GAM Worker (Algorithm~\ref{algo:worker}) threads\;
\KwRet{solutions}
\caption{P-GAM \label{algo:pgam}}
\end{algorithm}

\mysubsection{P-GAM: parallel keyword query execution}
\label{sec:parallelquery}

Our P-GAM (Parallel GAM) query algorithm  builds a set of data structures, which are exploited  by concurrent workers (threads) to produce query answers.
We split these data structures to shared and private to the workers.
We start with the shared ones.
The, \textbf{history} data structure holds all trees built during the exploration, while \textbf{treesByRoot} gives access to all trees rooted in a certain node.
% {\em These structures are shared by all the threads}, and the latter two are intensively read and written throughout excution; to enable maximum concurrency, we implemented them as lock-free hash-based maps.
As the search space is huge, the history and treesByRoot data structures grow very much.
Specfically, for history, P-GAM first has to make sure that an intermediate AT has not been considered before (i.e. browse the history) before writing a new entry.
Similar, treesByRoot is updated only when a tree changes its root or if there is a Merge of two trees; however, it is probed several times for Merge candidates.
Therefore, we have implemented these data structures as lock-free hash-based maps to ensure high concurrency and prioritize read accesses.
Observe that, given the high degree of data sharing, keeping these data structures thread-private would not yield any benefit.

Moving to the thread-private data structures, {\em each thread}, say number $i$, has a priority queue \textbf{pQueue$_i$}, in which are pushed (tree, edge) pairs, such that the edge is adjacent to the root of the tree. Priority in this queue is determined as follows: we prefer the pairs {\em whose nodes match most  query keywords}; to break a tie, we prefer {\em smaller trees}; and to break a possible tie among these, we prefer the pair where the edge has the {\em highest-specificity}. This is a simple priority order we chose empirically; any other priority could be used, with no change to the algorithm.

% \begin{figure}
%   \centering
%   \includegraphics[width=.85\columnwidth]{figures/pgam.png}
% \vspace{-6.5mm}
%   \caption{P-GAM runtime data structures.}
% \vspace{-7mm}
%   \label{fig:pgam}
% \end{figure}

P-GAM keyword search is outlined in Algorithm~\ref{algo:pgam}. It creates the shared structures,  and $nt$ threads (as many as available based on the availability of computing hardware resources). The search starts by looking up the nodes $N_Q$ matching at least one query keywords (line 2);  we create a 1-node tree from each such node, and push it together with an adjacent edge (line 4), in one of the pQueue's (distributing them in round-robin).

Next, $nt$ worker threads run in parallel Algorithm~\ref{algo:worker}, until a global stop condition: time-out, or until the maximum number of solutions has been reached, or all the queues are empty. Each worker repeatedly picks the highest-priority (tree, edge) pair on its queue (line 2), and applies Grow on it (line 3), leading to a 1-edge larger tree (e.g., $t_5$ obtained from $t_4$ in Figure~\ref{fig:GAM-trees}). Thus, the stack priority {\em orders} the possible Grow steps at a certain point during the search; it tends to lead to  small solutions being found first, so that users are not surprised by the lack of a connection they expected (and which usually involves few links).
If the Grow result tree had not been found before  (this is determined from the history), the worker tries to Merge it with all compatible trees, found within treesByRoot (line 6). The Merge partners (e.g., $t_5$ and $t_{15}$ in Figure~\ref{fig:GAM-trees}) should match different (disjoint) keywords; this condition ensures minimality of the solution. Merge results are repeatedly Merge'd again; the thread switches back to Grow only when no new Merge on the same root is possible. Any newly created tree is checked and, if it matches all query keywords, added to the solution set (and not pushed in any queue). Finally, to balance the load among the workers, if one has exhausted his queue, it retrieves the highest-priority (tree, edge) pair from the queue with most entries, pushing the possible results in its own queue.

\begin{algorithm}[t!]
\SetAlgoLined
\caption{P-GAM Worker (thread number $i$ out of $nt$)\label{algo:worker}}
\Repeat{{\em time-out} or $M$ {\em solutions are found} or {\em all pQueue$_j$ empty, for $1\leq j\leq nt$}}{
pop $(t,e)$, the highest-priority pair in pQueue$_i$ (or, if empty,  from the pQueue$_j$ having the most entries)\;
$t_G\leftarrow$ Grow($t,e$)\;
\If{$t_G\not\in$ history}%\nllabel{line:history-1}
   {for all edges $e'$ adjacent to the root of $t_G$, push ($t_G,e'$)  in pQueue$_i$\;
    build all $t_M\leftarrow$ Merge($t_G, t'$) where $t'\in$ treesByRoot.get($t_G$.root) and $t'$ matches $Q$ keywords disjoint from those of $t_G$\;\label{line:root-1}
    \If{$t_M \not\in$ history}%\label{line:history-2}
       {recursively merge $t_M$ with all suitable partners\; %\label{line:root-2}
        add all the  (new) Merge trees to history\; %\label{line:history-3}
        for each new Merge tree $t''$, and edge $e''$ adjacent to the root of $t''$, push $(t'',e'')$ in pQueue$_i$;
       }
  }
}
\end{algorithm}

As seen above,  the threads intensely compete for access to history %(lines \ref{line:root-1}, \ref{line:root-1} and \ref{line:root-3})
and treesByRoot. As we demonstrate in Section~\ref{sec:evalscalability}, our design allows excellent scalability as the number of threads increases.

\setlength{\textfloatsep}{.7\baselineskip}% Restore
\mysection{Experimental Evaluation}
\label{sec:evaluation}

We now present the results of our experimental evaluation.
Section~\ref{sec:evalsetup} presents the hardware and data used in our application.
Then, Section~\ref{sec:exp-loading} studies the impact of extraction policies (Section~\ref{sec:optims}).
Section~\ref{sec:evalscalability} analyzes the scalability of the P-GAM algorithm, focusing on its interaction with the hardware, and demonstrates its significant gains wrt GAM.
Section~\ref{sec:evalmacro} demonstrats P-GAM scalability on a large, real-world graph  built for our CoI IJ application.
%We split the evaluation in two parts: (i) scalability analysis, and (ii) absolute response time analysis.
%For the former, given in Section~\ref{sec:evalscalability}, we use synthetic data and we focus on the interaction of P-GAM with the hardware.
%For the latter, given in Section~\ref{sec:evalmacro}, we use real-world data that we have collected and ingested into our system and we discuss the impact of these times to the journalists.

\mysubsection{Hardware and Software Setup}
\label{sec:evalsetup}
We used a server with a 2x10-core Intel Xeon E5-2640 v4 (Broadwell) CPUs clocked at 2.4GHz, and 128GB of DRAM.
We do not use \href{https://en.wikipedia.org/wiki/Hyper-threading}{Hyper-Threads}, and we bind every CPU core to a single worker thread.
As shown in Figure~\ref{fig:archi}, ConnectionLens (90\% Java, 10\% Python) is used (Section~\ref{sec:exp-loading}) to {\em construct} a graph out of a set of data sources, and {\em store} it in PostgreSQL. Next in the processing pipeline, we {\em migrate} the graph to
%We use different software for ingesting the data and building the graph and different software for querying the graph.
%Data ingestion has been developed in Python and includes modules for entity extraction from raw data sources.
%The extraction results are passed to the ConnectionLens application, implemented in Java, which models them as a graph and stores it in PostgreSQL for portability across different applications that we develop in the course of the project.
%Observe that the process described so far takes place only when there are new data to ingest to the system.
the novel {\em in-memory graph engine} previously describe, which queries it using the P-GAM algorithm.
%Instead, for querying, we have developed an in-memory graph query engine implementing P-GAM that we describe in this paper.
The query engine is a NUMA-aware, multi-threaded C++ application.

\mysubsection{The Impact of Extraction Policies}
\label{sec:exp-loading}
In this experiment, we loaded a set of $20.000$  bibligraphic Pubmed XML bibliographic notices ($38.4$ MB on disk). This dataset inspired an extraction policy stating that: the text content of any PubMedArticle.Authors.Author.Name is a Person entity, and that extraction is skipped from the article and journal title, as well as from the article keywords. NER is still applied on author affiliations (rich with Organization and Location entities), as well as on the CoIStatement elements of crucial interest in our context.

\begin{table}
\begin{tabular}{|l|r|r|r|}
\hline
          & Total  (s) & Extraction (s) & Storage (s) \\
\hline\hline
\textbf{No policy} &1416& 1199 & 136 \\
\hline
\textbf{Using policy} &  929 & 716 & 131\\
\hline
\end{tabular}
\vspace{1mm}
\caption{Sample impact of an extraction policy.\label{tab:policy}}
\vspace{-8mm}
\end{table}

Table~\ref{tab:policy} shows that our policy reduced the extraction time by about 40\%,  reducing the loading time by 34\%. As a point of reference, we also noted the time to load (and index) the graph nodes and edges in PostgreSQL; extraction strongly dominates the total time, confirming the practical interest of application-driven policies.

\mysubsection{Scalability Analysis}
\label{sec:evalscalability}

The scalability analysis is performed on synthetic graphs, whose size and topology we can fully control.
%Given that P-GAM is exhaustive and is used for interactive exploration of graphs which have been created in a completely ad-hoc manner, we do not make any assumptions on its properties and, therefore, we devise no graph-specific heuristics to reduce the search space.
We focus on two aspects that impact scalability: ($i$)~contention in concurrent access to %stress on concurrent
data structures, and ($ii$)~size of the graph (which impacts the search space).
To analyze the behavior of P-GAM's concurrent data structures, we use  chain$_k$ graphs, because they yield a large number of intermediate results, shared across threads, even for a small graph.
This way, we can isolate the size of the graph from the size of the intermediate results.

\begin{figure}
\includegraphics[width=.55\columnwidth]{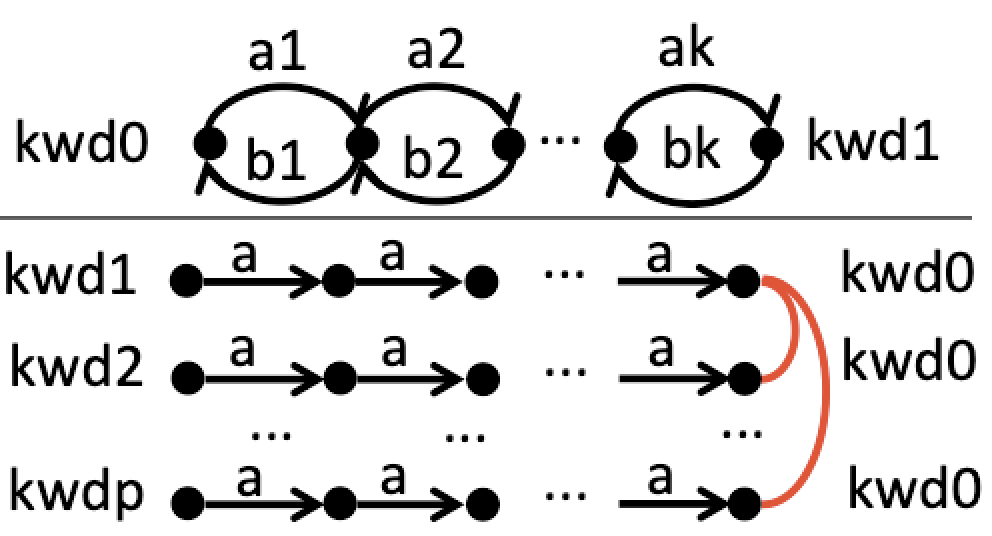}
\vspace{-1.5mm}
\caption{Synthetic graphs: chain$_k$ and star$_{p,k}$.\label{fig:synthetic}}
\vspace{-3mm}
\end{figure}

We use \textbf{two shapes of graphs (each with 1 associated query)},  leading to very different search space sizes (Figure~\ref{fig:synthetic}). In both graphs, all the kwd$_i$ for $0\leq i$ are distinct keywords, as well as the labels of the node(s) where the keyword is shown; no other node label matches these keywords.
\textbf{Chain$_k$} has $2k$ edges; on it,  \{kwd$_0$, kwd$_1$\} has $2^k$ solutions, since any two neighbor nodes can be connected by an $a_i$ or by a $b_i$ edge; further, $2^{k+1}-2$ partial (non-solution) trees are built, each containing one keyword plus a path growing toward (but not reaching) the other.
%\IM{Yamen and I agree: from kwd$_0$ we have 2 paths reaching the closest node, 4 reaching the next one, ... $2^{k-1}$ reaching the node just before kwd$_0$, thus $\sum_{i=0}^{k-1}(2^i)=2^k-1$. Since a symetric set of trees start from kwd$_1$ and go towards kwd$_0$, we obtain $2\cdot(2^k-1)=2^{k+1}-2$.}
\textbf{Star$_{p,k}$} has $p$ branches, each of which is a line of length $k$; at one extremity each line has a keyword kwd$_i$, $1\leq i \leq p$, while at the other extremity, all lines have kwd$_0$. As explained in Section~\ref{sec:build-graph}, these nodes are equivalent, one is designated their representative (in the Figure, the topmost one), and the others are connected to it through equivalence edges, shown in red. On this graph, the query \{kwd$_0$, kwd$_1,\ldots$, kwd$_p$\} has exactly 1 solution which is the complete graph; there are $O(k+1)2^p$ partial trees. %\IM{Deduction worked out with Yamen: We have
\begin{table}
\scalebox{0.9}{\begin{tabular}{|l||r|r|r|r|r|}
\hline
Graph                & $S$ & $T^1_{PGAM}$ (ms) & $T_{PGAM}$ (s)  & $T^1$ (ms) & $T$ (s)\\
\hline\hline
chain$_{12}$	 &4096      &2	      &	0.8	&	 160 & 674.5\\ % IM: CL Updated 31/1
chain$_{13}$     &8192      &4          &	  3.8         & 203 & 900.0\\ % IM: CL Updated 31/1
chain$_{14}$	&16384     & 4	&	        13.7&	     234 & 900.0\\ % IM: CL Updated 31/1
chain$_{15}$	&32768     & 8	&		53.2&	   315  & 900.0\\ % IM: CL Updated 31/1
star$_{4,1000}$	&        1     &	233        &	.6&	        4063  & 60.2 \\ % IM: CL Updated 31/1
star$_{4,2000}$	&1             & 969	        &     3.3 &    12580  & 243.9\\
star$_{4,3000}$	&1             & 2469	        &     10.1&    36261  & 900.0 \\
star$_{4,4000}$	&1             &	5149        &	23.3&    67984   & 900.0\\
star$_{4,5000}$ &1             & 9111	        &	44.7	& 108960& 900.0\\
\hline
\end{tabular}}
\vspace{1mm}
\caption{Single-thread P-GAM vs. GAM performance.\label{fig:gam-compare}}
\vspace{-7mm}%\caption{title}
\end{table}
% \end{figure}

\noindent\textbf{Single-thread P-GAM vs. GAM} We start by comparing P-GAM, {\em using only 1 thread}, with the (single-threaded) Java-based GAM, accessing graph edges from a PostgreSQL database. We ran the two algorithms on the synthetic graphs and queries, with a {\em time-out of 15 minutes}; both could stop earlier if they exhausted the search space. Table~\ref{fig:gam-compare} shows: the number of solutions $S$, the time $T^1_{PGAM}$ (ms) until the first solution is found by PGAM and its total running time $T_{PGAM}$ (s), as well as the corresponding times $T^1$ and $T$ for GAM (Java on Postgres). On these tiny graphs, both algorithms found all the expected solutions, however, even without parallelism, P-GAM is  $10\times$ to more than $100\times$ faster. In particular, on all but the 3 smallest graphs, GAM did not exhaust its search space in 15 minutes. This experiment validates the expected orders of magnitude speed-up of a carefully designed in-memory implementation, even without parallelism (since we restricted P-GAM to 1 thread).

\begin{figure}
\centering
\includegraphics[width=\columnwidth]{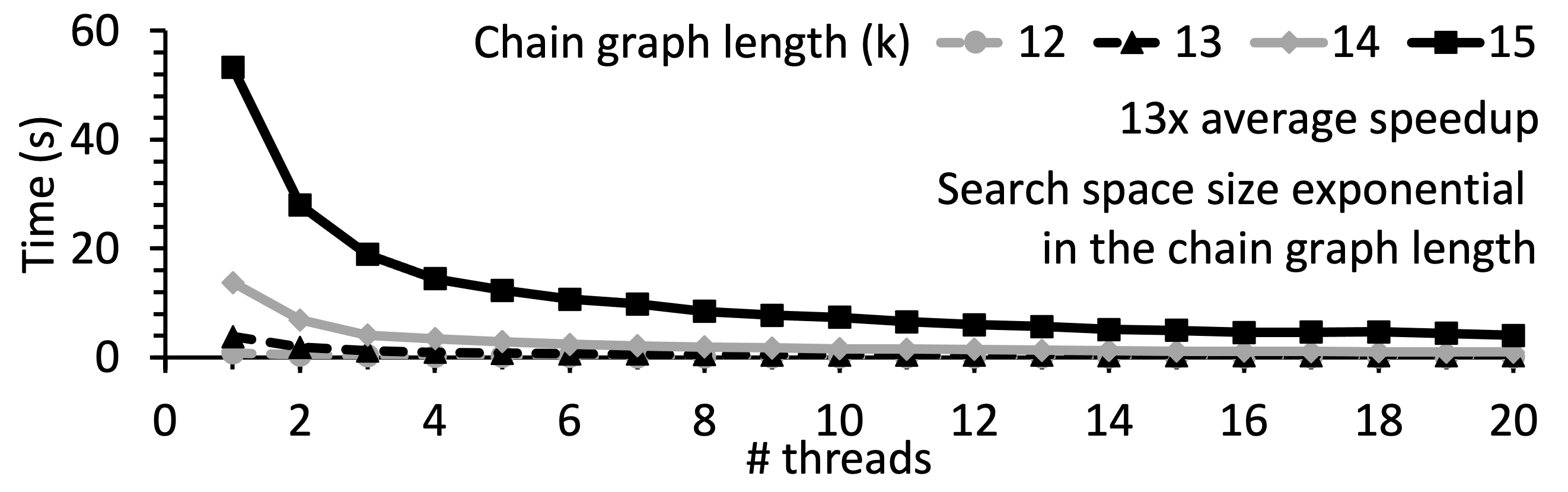}
\vspace{-6mm}
\caption{GAM-P scaling on chain graphs.}
\label{fig:evalchain}
\vspace{.5mm}
\end{figure}

\noindent\textbf{Parallel P-GAM} Next, on the graphs chain$_k$ for $12\leq k \leq 15$, we report the exhaustive search time (Figure~\ref{fig:evalchain}) for query \{kwd$_0$, kwd$_1$\}  as we increase the number of worker threads from 1 to 20. We see a clear speedup as the number of threads increases, which is on average 13x for the graph sizes that we report.
The speedup is not linear, because as the size of the intermediate results grows, it exceeds the size of the CPU caches, while threads need to access them at every iteration.
Our profiling revealed that, as several threads access the shared data structures, they evict content from the CPU cache that would be useful to other threads.
Instead, we did not notice overheads from our synchronization mechanisms.
% Profiling our system showed that when the number of threads increases,  the memory access requests  overlap, and cache locality is improved; data structures are more likely to be in the processor's L3 cache.
% Given that the most expensive operations are the ones that check whether an AT has been examined before, bringing the hash buckets of the history map in the CPU caches improves locality. This is why \textbf{the speed-up is actually higher than the number of threads}.

\begin{figure}
\centering
\includegraphics[width=\columnwidth]{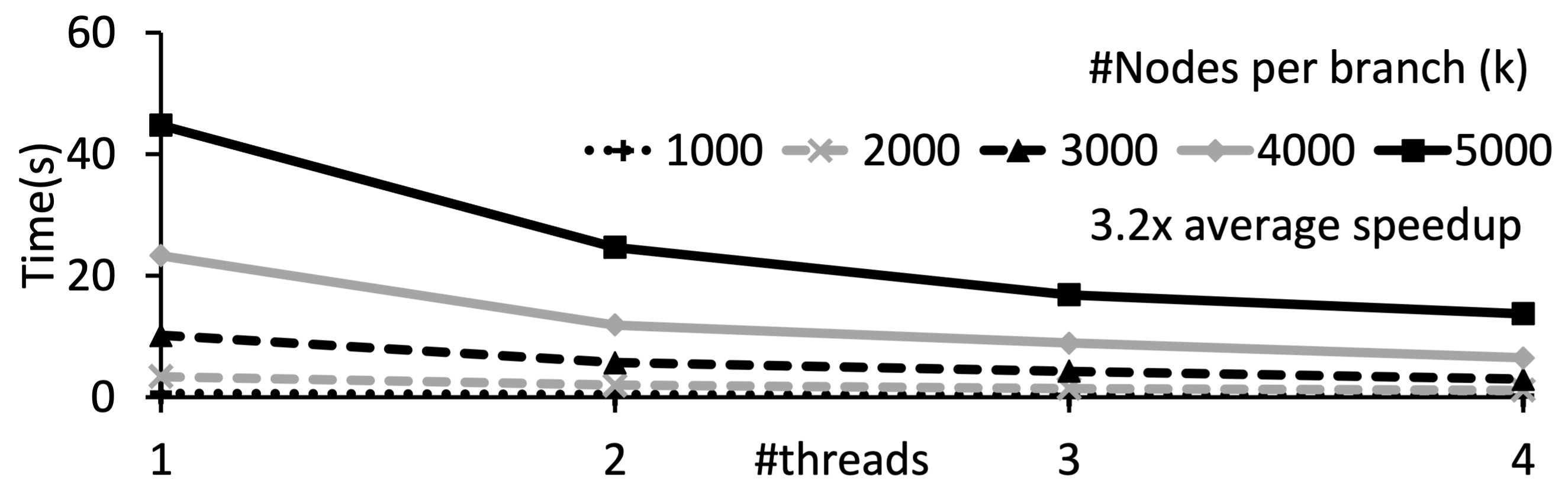}
\vspace{-6mm}
\caption{P-GAM scaling on star graphs.}
\vspace{-2.5mm}
\label{fig:evalstar}
\end{figure}

To study the scalability of the algorithm with the graph size, we use $star_{4,k}$ for $k\in \{1000, 2000, 3000, 4000, 5000\}$ and the query \{kwd$_1$, kwd$_2$, kwd$_3$, kwd$_4$\}.
Figure~\ref{fig:evalstar} shows the exhaustive search time of P-GAM on these graphs of up to $20.000$ nodes, using $1$ to $4$ threads.
We obtain an average speed-up of $3.2\times$ with $4$ threads, regardless the size of the graph, which shows that P-GAM scales well for different graph models and graph sizes.
After profiling, we observed that the size of the intermediate results impacts the performance, similar to the previous case of the chain graph.

In the above star$_{4,k}$ experiments, we used up to $4$ threads since the graph has a symetry of $4$ (however, theads share the work  with no knowledge of the graph structure). When keyword matches are poorly connected, e.g., at the end of  simple paths, as in our star graphs, P-GAM search starts by exploring these paths, moving farther away from each keyword; if $N$ nodes match query keywords, up to $N$ threads can share this work. In contrast, as soon as these explored paths intersect, Grow and Merge create many opportunities that can be exploited by one thread or another.
% \IM{I have issues  with what follows. I tried to rewrite its idea above.\\}
% We have seen that in graph topologies which consist of lines (single or more) where nodes are connected with a single edge, \IM{I have a hard time understanding the topologies meant here. Is it ``in star$_{p,k}$ graphs''? If yes, we should say so. If not, it may be hard for users to follow exactly what the topologies are.}
% P-GAM scales up to the number of keyword-matching nodes, \IM{If there are $4$ keywords and $333$ nodes matching them ($80$, $80$, $80$, and $3$, respectively), do we mean P-GAM scales well up to $4$ or $333$ threads?} and in this case four.
% This is explained by the operations of the algorithm:
% P-GAM starts from the nodes which match the keywords and it applies a Grow operation (i.e. expans the AT with one edge) until another AT with the same root is found.
% At that point, it will perform a Merge operation to merge the two trees. \IM{Ok with the previous two phrases.}
% Therefore, there is only a single operation which can be applied every time. \IM{Not really: suppose 3-kwd query matched in 3 branches such that the kwds are at distance 100, 50, and 10 from the ``central root'' that connects them. We start from the 3 kwds. At any point we can do one of 3 Grow steps. After 10 Grow, we can keep growing in the ``neighbor'' branches, after the central node... So, I have a hard time with {\em only a single operation can be applied every time}.}
On chain$_k$, the presence of 2 edges between any adjacent nodes multiplies the Grow and Merge opportunities, work which can be shared by many threads.
This is why on chain$_k$, we see scalability up to 32 worker threads, which is the maximum that our server supports.

% \noindent\textbf{Note on intermediate results.}
% We have observed that the size of intermediate results (millions of trees) are limiting scalability.
% However, in practice, we have seen that trees which are very big or similar are not of interest.
% Therefore, various heuristics can be applied to prune the search space and avoid these bottlenecks without necessarily harming the quality of the results for the end-users.
% Nevertheless, here we gave the full picture for completeness.

\mysubsection{P-GAM in Conflict of Interest Application}
\label{sec:evalmacro}

We now describe  experiments on actual application data.

\noindent\textbf{The graph.} We selected sources based on S.~Horel's expertise and suggestions, as follows. ($i$)~We loaded \textbf{400.000} PubMed  bibliographic notices (\textbf{XML}), corresponding to articles from 2019 and 2020; they occupy \textbf{803 MB} on disk. We used the  same extraction policy as  in Table~\ref{tab:policy} to perform only the necessary extraction.
%% \IM{We have 412 files loaded, each is about 1.95 MB.}
($ii$)~We have downloaded \textbf{85.400} PDF articles corresponding to these notices (those that were available in Open Access),
%cl_pubmed_build_2020=> select count(*) from edges e, catalog c where e.ds=c.id and c.type='JSON' and e.label='PDF_pmid_URI';
% count
%-------
% 85418
transformed them into \textbf{JSON} using an extraction script we developed,
%building upon that described in~\cite{DBLP:journals/corr/abs-2012-08830}
and preserved only those paragraphs starting with a set of keywords (``Disclosure'', ``Competing Interest'', ``Acknowlegments'' etc.) which have been shown~\cite{monsanto}  to encode potentially interesting participations  of people (other than authors) and organizations in an article. Together, these JSON fragments occupy \textbf{173 MB} on disk.
% [manolesc@cedar005 bouganim]$ ls -l ./pubmed/pubmed-pipeline/json_extract_01_29_2021/json_extract_of_1000_18.json
% -rw-r--r--. 1 tbougani cedar 2112753 29 janv. 17:05 ./pubmed/pubmed-pipeline/json_extract_01_29_2021/json_extract_of_1000_18.json
% [manolesc@cedar005 bouganim]$ bc -l
% bc 1.06.95
% 2112753*86/(1024*1024)
% 173.27953147888183593750
{\em The JSON and the XML content from the same paper are connected (at least) through  the URI of that paper, as shown in Figure~\ref{fig:motiv}.}
($iii$)~We have crawled 375 \textbf{HTML} Web pages from a set of Web sites describing people and organizations previously involved in scientific expertise on sensitive topics (such as tobacco or endocrine disruptors), specifically: www.desmogblog.com, tobaccotactics.org, www.wikicorporates.org and www.sourcewatch.org. These pages total \textbf{31.97 MB}.
% find . -name \*.html | xargs wc
% 33531556 total
% [mac9:trunk/PAPERS/CLMem--VLDB21-SDS] ioanamanolescu% bc -l
% 33531556/(1024*1024)
% 31.97818374633789062500
Table~\ref{tab:graph-stats} shows the numbers of nodes $|N|$, of edges $|E|$, and, respectively, of Person, Organization and Location entities ($|N_P|$, $|N_O|$, $|N_L|$), split by the data model, and overall.

\begin{table}
\centering
\scalebox{0.95}{\begin{tabular}{|l||r|r|r|r|r|}
\hline
$|N|$   & $|E|$       & $|N|$         & $|N_P|$       & $|N_O|$  & $|N_L|$   \\
\hline\hline
XML   & 32,028,429 &19,851,904 &   1,483,631&584,734  & 126,629 \\
\hline
JSON &1,025,307    &432,303     &     75,297  &  7,320    &    4,139 \\
\hline
HTML &  246,636   & 185,479    &      3,726    &    7,227  &    320   \\
\hline
Total &  33,300,372 & 20,469,686& 1,562,654   &  665,167 & 131,088 \\
\hline
\end{tabular}}
\caption{Statistics on Conflict of Interest application graph.\label{tab:graph-stats}}
\vspace{-8mm}
\end{table}

\begin{table}
  \scalebox{0.95}{
    \begin{tabular}{|c|l||r|r|r|r|r|}
      \hline
      $\#$ & Keywords  & $T^1$ & $T^{last}$ & $T$ & $S$ & \# $DS$ \\
      \hline\hline
      1 & A1, A2 & 4462 & 5315 & 5316 & 1000 & 2-10, \textbf{\underline{6}} \\
      \hline
      2 & A3, H1 & 4671 & 5140 & 5140 & 1000 & 3-7, \textbf{\underline{6}} \\
      \hline
      3 & U1, H1 & 4832 & 4981 & 4981 & 1000 & 2-5, \textbf{\underline{5}} \\
      \hline
      4 & A4, I1 & 8520 & 13711 & 13712 & 1000 & 2-5, \textbf{\underline{5}} \\
      \hline
      5 & A5, I2 & 5800 & 6366 & 6366 & 1000 & 2-8, \textbf{\underline{8}} \\
      \hline
      6 & A6, I3, P1 & 4657 & 5072 & 60000 & 16 & 4, \textbf{\underline{4}} \\
      \hline
      7 & A7, I3, P2 & 44256 & 44273 & 60000 & 10 & 5, \textbf{\underline{5}} \\
      \hline
      8 & A8, I4, P3 & 12560 & 12560 & 60000 & 2 & 5, \textbf{\underline{5}} \\
      \hline
      9 & A9, I4, P3 & 28982 & 33435 & 60000 & 3 & 5, \textbf{\underline{5}} \\
      \hline
      10 & A10, U1, I3 & 7577 & 17383 & 17383 & 1000 & 4-6, \textbf{\underline{6}} \\
      \hline
      11 & A11, I4, I5 & 10396 & 32320 & 60000 & 6 & 3, \textbf{\underline{3}} \\
      \hline
      12 & A12, I4, I6 & 7320 & 7467 & 60000 & 24 & 4, \textbf{\underline{4}} \\
      \hline
      13 & A3, A13, U2, P4 & 15759 & 35025 & 60000 & 5 & 5-6, 8, \textbf{\underline{6,8}} \\
      \hline
      14 & A3, A14, U3, G1 & 10711 & 10711 & 60000 & 1 & 7, \textbf{\underline{7}} \\
      \hline
      15 & A3, A15, U4, P4 & 8560 & 9942 & 60000 & 16 & 9, \textbf{\underline{9}} \\
      \hline
    \end{tabular}
  }
  \caption{P-GAM performance on CoI  real-world graph.\label{tab:realworld}}
  \vspace{-4mm}
\end{table}

\noindent\textbf{Querying the graph.} Table~\ref{tab:realworld} shows the results of executing 15 queries, until \textbf{1000 solutions} or for at most \textbf{1 minute}, using P-GAM.
From left to right, the columns  show: the query number, the query keywords, the time $T^1$ until the first solution is found, the time $T^{last}$ until the last solution is found, the total running time $T$,  the number of solutions found, and some statistics on the number of data sources participating in the solutions found ($\#DS$, see below). All times are in milliseconds.
We have anonymized the keywords that we use, not to single out individuals or corporations, and since the queries are selected aiming not at them, but at a large variety of P-GAM behavior.
%since the objective of this work is to demonstrate the effectiveness of our approach in discovering results in interactive time.
We use the following codes: \textbf{A} for author,  \textbf{G} for government service, \textbf{H} for hospital, \textbf{P} for country, \textbf{U} for university, and \textbf{I} for industry (company).
A $\#DS$ value of the form ``2-10, \textbf{\underline{6}}'' means that P-GAM found solutions spanning at least 2 and at most 10 data sources, while most solutions spanned over 6 sources.

We make several observations based on the results.
%First, we note that we set our query engine to run for maximum one minute or until it finds 1000 solutions, explaining why in all queries we either have $T_{total} = 60000ms$ or $Sol=1000$.
The stop conditions were set  here based on what we consider as an interactive query response time, and a number of solutions which allow further exploration by the users (e.g., through an interactive GUI we developed).
Further, solutions span over several datasets, demonstrating the interest of multi-dataset search enabled, and that P-GAM exploits this possibility.
Finally, we report results after performing queries including different amount of keywords and the system remains responsive within the same time bounds, despite the increasing query complexity.

\mysection{Related Work and Conclusion}
\label{sec:related}

In this paper, we presented a complete pipeline for managing heterogeneous data for IJ applications. 
%Our pipeline includes the steps which are required for both ingestion and querying of the data, whereas we present our contributions in both aspects.
This innovates upon recent work~\cite{DBLP:journals/corr/abs-2012-08830} where we have addressed the problems of integrating such data in a graph and querying it, as follows: ($i$)~we present a complete data science application with clear societal impact, ($ii$)~we show how extraction policies  improve the graph construction performance, and ($iii$)~we introduce a parallel search algorithm which scales across different graph models and sizes.
Below, we discuss prior work most relevant wrt the contributions we made here; more elements of comparison can be found in~\cite{DBLP:journals/corr/abs-2012-08830}. 

Our work falls into the {\em data integration} area~\cite{DBLP:books/daglib/0029346}; our IJ pipeline starts by ingesting data into an integrated data repository, deployed in PostgreSQL. The first platform we proposed to Le Monde journalists was a mediator~\cite{DBLP:journals/pvldb/BonaqueCCGLMMRT16}, resembling polystores, e.g.,~\cite{DugganESBHKMMMZ15,DBLP:journals/dpd/KolevVBJPP16}. However, we found that: ($i$)~their datasets are changing, text-rich and schema-less, ($ii$)~running a set of data stores (plus a mediator) was not feasible for them, ($iii$)~knowledge of a schema or the capacity to devise integration plan was lacking. ConnectionLens' first iteration~\cite{DBLP:journals/pvldb/ChanialDGLNM18} lifted ($iii$)~by introducing keyword search, but it still kept part of the graph {\em virtual}, and split keyword queries into subqueries sent to sources. Consolidating the graph in a single store, and the centralized GAM algorithm~\cite{DBLP:journals/corr/abs-2012-08830} greatly sped up and simplified the tool, whose performance we again improve here. 
We share the goal of exploring and connecting data,  with  {\em data discovery}  methods~\cite{DBLP:conf/sigmod/SarmaFGHLWXY12,DBLP:conf/icde/FernandezAKYMS18,DBLP:conf/icde/FernandezMQEIMO18,DBLP:journals/pvldb/OtaMFS20}, which have mostly focused on tabular data. While our data is heterogeneous, focusing on an IJ application partially eliminates risks of ambiguity, since in our context, one person or organization name typically denote a single concept. 

{\em Keyword search} has been studied in XML~\cite{DBLP:conf/sigmod/GuoSBS03, DBLP:conf/sigmod/LiuC07}, graphs (from where we borrowed Grow and Merge operations for GAM)~\cite{DBLP:conf/icde/DingYWQZL07, DBLP:conf/sigmod/HeWYY07}, and in particular RDF graphs~\cite{DBLP:conf/cikm/ElbassuoniB11, DBLP:journals/tkde/LeLKD14}.
However, our keyword search problem is harder in several aspects: 
($i$)~we make no assumption on the shape and regularity of the graph; ($ii$)~we allow answer trees to explore edges in both directions; 
($ii$)~we make no assumption  on the score function, invalidating Dynamic Programming (DP) methods such as~\cite{DBLP:conf/sigmod/LiuC07} and other similar prunings. In particular, we show in~\cite{DBLP:journals/corr/abs-2009-04283} that {\em edges with a confidence lower than 1}, such as similarity and extraction edges in our graphs, compromise, for  any ``reasonable'' score function which reflects these confidences,  the {\em optimal substructure} property at the core of DP.  
Works on {\em parallel keyword search in graphs} either consider a different setting, returning a certain class of subgraphs instead of trees~\cite{DBLP:conf/icde/YangAJTW19} or standard graph traversal algorithms like BFS~\cite{DBLP:conf/IEEEpact/HongOO11, DBLP:conf/spaa/DhulipalaBS17, 10.1145/1810479.1810534}.
To the best of our knowledge, GAM is the first keyword search algorithm for the specific problem that we consider in this paper.
Accordingly, in this paper we have parallelized GAM, into P-GAM, by drawing inspiration and addressing common challenges raised in graph processing systems in the literature, in particular concerning the CPU efficiency while interacting with the main memory~\cite{DBLP:conf/usenix/MalicevicLZ17, DBLP:conf/isca/AhnHYMC15, DBLP:conf/fast/ElyasiCS19, DBLP:conf/sosp/RoyMZ13, DBLP:conf/sc/HongDMLVC15}.

Our future work  includes: building a unified CoI repository based on more biomedical sources, enhancing our in-memory query processor, and querying the graph using natural language. 
%\input{conclusion}

% \begin{acks}
%  This work was supported by the [...] Research Fund of [...] (Number [...]). Additional funding was provided by [...] and [...]. We also thank [...] for contributing [...].
% \end{acks}
\vspace{1.5mm}
\noindent\textbf{Acknowledgments.} The authors thank M.~Ferrer and
the Décodeurs  team (Le Monde) for
introducing us, and for many insightful
discussions.
% \clearpage

\bibliographystyle{ACM-Reference-Format}
\bibliography{main}

\end{document}